\documentclass[sigconf]{acmart}

\usepackage{algorithmic}
\usepackage{graphicx}
\usepackage{textcomp}
\usepackage{xcolor}
\usepackage{fontawesome5}
\usepackage{booktabs}
\usepackage{multirow}
\usepackage{tabularx}
\usepackage{url}
\usepackage{subcaption}
\usepackage{float}

\usepackage{tikz}
\usetikzlibrary{decorations.pathreplacing, patterns, positioning, calc, backgrounds, patterns, calligraphy}

\AtBeginDocument{%
  }

\setcopyright{cc}
\setcctype{by}
\copyrightyear{2026}
\acmYear{2026}
\acmDOI{10.1145/3846375.3849104}
\acmConference[WTMC '26]{11th Workshop on Traffic Measurements for Cybersecurity}{November 15--19, 2026}{The Hague, Netherlands}
\acmBooktitle{11th Workshop on Traffic Measurements for Cybersecurity (WTMC '26), November 15--19, 2026, The Hague, Netherlands}
\acmISBN{979-8-4007-3017-7/2026/11}

\begin{document}

\title{Learning to Link: Automatic Re-identification of BLE Devices Under MAC Address Randomisation}


\author{Reem Abdulrhman Alghamdi}
\affiliation{%
  \institution{King Abdullah University of Science and Technology (KAUST)}
  \city{Thuwal}
  \country{Saudi Arabia}}
\email{reem.alghamdi.1@kaust.edu.sa}

\author{Alberto Verna}
\affiliation{%
  \institution{Politecnico di Torino}
  \city{Torino}
  \country{Italy}
}
\email{alberto.verna@polito.it}

\author{Marco Mellia}
\affiliation{%
 \institution{Politecnico di Torino}
 \city{Torino}
 \country{Italy}}
\email{marco.mellia@polito.it}

\renewcommand{\shortauthors}{Alghamdi et al.}

\begin{abstract}
Bluetooth Low Energy (BLE) employs MAC address randomisation -- via Resolvable Private Address (RPA) -- to mitigate long-term device tracking on public advertising channels. Existing research has shown that advertising packets contain metadata and structural features that allow re-identifying a target device via manually crafted rules.
In this work, we investigate the feasibility of automating the process of tracking BLE devices despite MAC randomisation by leveraging machine learning algorithms for the signature creation.
Based on the actual Bluetooth traffic from target devices, we characterise the persistence of advertising-layer features across RPA changes and formulate device linkage as a supervised classification problem. Using simple decision tree classifiers as a proof-of-feasibility approach, we evaluate the distinguishability of target and non-target devices under varying address rotation patterns.
Our results reinforce prior work demonstrating that advertising-layer metadata can enable device re-identification under MAC randomisation, to the point where such linkage can be automated using standard supervised learning techniques, without any specific knowledge of the technology.
\end{abstract}




\maketitle

\section{Introduction}
\label{sec:Introduction}
Bluetooth Low Energy (BLE) is the most widely used standard for short-range wireless communication between personal devices, wearables, and IoT systems.
Back in 2010, the BLE 4.0  standard introduced the Resolvable Private Address (RPA) standard to mitigate device tracking and protect user privacy. A device uses \textit{randomised MAC addresses} instead of its real MAC address, i.e., it periodically refreshes its address to prevent observers from long-term device tracking~\cite{b22, b23}. Only paired devices that possess the key to resolve the randomised address can identify the real device MAC and use its services.


Despite its design goals, prior research has shown that BLE address randomisation does not fully prevent device tracking. Several works show that randomised addresses can be linked to the same physical device, either by exploiting implementation flaws \cite{b2, b4}, side channels \cite{b7, b8} or information leakage from payload data \cite{b9, b10, b13, b15, b16, b18}. 
Existing attacks, however, still require substantial protocol knowledge and manual analysis. In particular, effective tracking often depends on handcrafted heuristics, expert interpretation of payload structure, or device-specific signature design. This limits the scalability of the attack and raises the bar for the attacker.


In this paper, we ask a different question: Does modern machine learning make BLE de-randomisation easy to automate? More precisely, we investigate whether a passive adversary can train a classifier from previously observed advertising traffic and later use it to recognise the presence of the same target device after one or more MAC address rotations. Our goal is not to introduce a stronger side channel, but to show that advertising-layer information already observable by a passive receiver is sufficient to automate the attack.

We formulate BLE device re-identification as a supervised binary classification problem over advertising packets. In an initial training phase, the adversary observes the target device under conditions that allow its advertisements to be labelled as positive samples, while also collecting advertising packets from unrelated nearby devices as negative samples. From these labelled packets, the adversary trains a model that captures advertising-layer characteristics observable without pairing or address resolution. During deployment, the trained classifier identifies packets likely emitted by the same target device in a mixed BLE trace. Packet-level predictions are then aggregated by advertised MAC address to determine whether a previously unseen RPA is a candidate address of the target device. This shifts the attack from manual, device-specific signature construction to automated feature learning.
We further compare this approach against exact matching of previously observed advertising payloads, allowing us to quantify the benefit of learning persistent payload patterns rather than relying on complete payload invariance.


We evaluate the classification framework on traces collected from {fourteen} target devices, aiming to re-identify each within a pool of several hundred devices under varying traffic conditions and noise levels. In our dataset, we surprisingly observe that, 15 years after the introduction of Resolvable Private Addresses, only 1/3 of observed devices do not even support it.
Even when trained on only two observed RPA epochs, a simple decision tree can recover almost all advertisements from the target (98.6\% recall), with a packet-level FPR below 2\% and only 0.38\% false device identifications after aggregation. These results reconfirm established work in the field that packet-level features observable on BLE advertising channels remain sufficiently stable across MAC address changes to support automatic re-identification. More importantly, they show that machine learning substantially reduces the expertise needed to mount this class of attack, enabling automated and scalable device tracking.






\section{Bluetooth Low Energy}
\label{sec:ble}
This section provides background on Bluetooth Low Energy (BLE), focusing on the advertising mechanism and device-addressing model on which our work builds.


\subsection{BLE Protocol}
\label{subsec:ble-protocol}

BLE is a short-range communication technology operating on 40 channels in the 2.4 GHz ISM band~\cite{b23}. Three of these channels are reserved for \emph{advertising} traffic, while the remaining are used for data transfer.

BLE devices typically operate in a connectionless discovery mode before establishing encrypted communications. Devices periodically broadcast advertising packets that can be passively received by any nearby observer, enabling device discovery and announcing available services.

\subsection{BLE Device Advertising}
\label{subsec:ble-advertising}

Bluetooth devices announce their presence and capabilities through advertising packets without requiring prior pairing. In this work, we focus on BLE Extended Advertising (EA) packets, which contain the advertised device address and other protocol-level fields.
These fields include Advertising Data (AD), where devices may disclose information such as names, services, capabilities, and manufacturer-specific data.

All advertising-layer fields are transmitted in plain text before connection establishment. Therefore, both header metadata and advertising data structures are observable by a passive adversary.

\subsection{BLE Device Addressing}
\label{subsubsec:ble-addressing}

BLE devices are identified by a 48-bit Bluetooth MAC address, which can be of multiple address types \cite{b23}:
\begin{itemize}
    \item \textbf{Public Device Address (PDA):} an address uniquely assigned to the device by the manufacturer in compliance with IEEE specifications \cite{ieee_ra}.

    \item \textbf{Random Static Address (RSA):} a randomly generated address that
    remains unchanged during device operation and is only
    renewed after a power cycle.

    \item \textbf{Non-Resolvable Private Address (NRPA):} a randomly generated address
    that is periodically renewed and cannot be resolved to a permanent
    identity.

    \item \textbf{Resolvable Private Address (RPA):} a random address that can be resolved to a stable identity using an Identity Resolution Key (IRK). Like NRPAs, this class of address is refreshed periodically, typically every 15 minutes.
\end{itemize}

MAC addresses are present in all advertising messages. The main goal of RPAs is to prevent long-term linkability of advertising packets to a single physical device based solely on the observability of the device's MAC address. However, address rotation only protects the MAC field, while the content and structure of advertising data are implementation-dependent and not strictly bound to the address regeneration process.

\subsection{Threat Model}
\label{sec:threat-model}

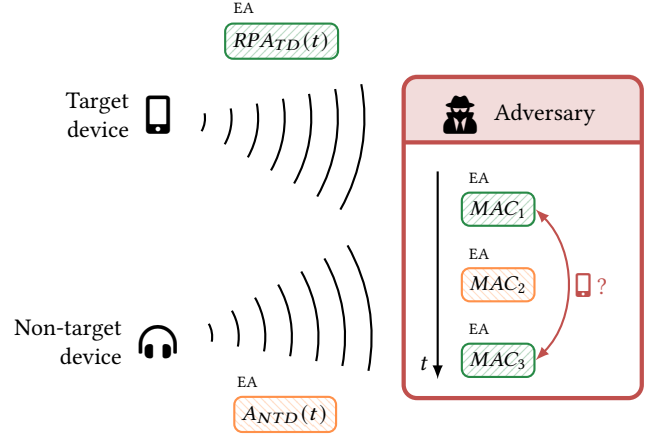
\begin{figure}
    \centering
    \pgfdeclarelayer{bg}
\pgfsetlayers{bg,main}

\definecolor{adversary}{HTML}{C0504D}
\definecolor{target}{HTML}{288A47}
\definecolor{target-bg}{HTML}{BEDBC7}
\definecolor{nontarget}{HTML}{FF914A}
\definecolor{nontarget-bg}{HTML}{FFDEC8}
\definecolor{anon1}{HTML}{03468F}
\definecolor{anon1-bg}{HTML}{cee5fe}
\definecolor{anon2}{HTML}{D20000}
\definecolor{anon2-bg}{HTML}{ffd2d2}
\definecolor{anon3}{HTML}{82218b}
\definecolor{anon3-bg}{HTML}{e6d2e7}

\tikzset{
    packet/.style={
        draw,
        thick,
        font=\small,
        rounded corners=3pt,
        align=left
    },
    target/.style = {
        draw = target,
        pattern = north east lines,
        pattern color = target-bg
    },
    nontarget/.style = {
        draw = nontarget,
        pattern = north west lines,
        pattern color = nontarget-bg
    }
}

\begin{tikzpicture}

    \node[label={[align=right]left:Target\\device}] (target) at (0,0) {\huge \faIcon{mobile-alt}};
    \node[label={[align=right]left:Non-target\\device}] (non-target) at (0,-3) {\huge \faIcon{headphones-alt}};

    \draw[decoration={expanding waves,angle=20},decorate, thick]
        (target)
        -- node[packet, target, above=1cm, label={above left:{\scriptsize EA}}] {$RPA_{TD}(t)$}
        (3,-.5);
    \draw[decoration={expanding waves,angle=20},decorate, thick]
        (non-target)
        -- node[packet, nontarget, below=1cm, label={above left:{\scriptsize EA}}] {$A_{NTD}(t)$}
        (3,-2.5);
    
    \node[label={right:Adversary}] (adversary) at (4,0) {\huge \faIcon{user-secret}};

    \draw[-latex, thick] (3.7, -.75) -- (3.7, -3.5) node[above left] {$t$};
    
    \node[packet, target] (mac1) at (4.5,-1.25) {$MAC_1$};
    \node[packet, nontarget] (mac2) at (4.5,-2.25) {$MAC_2$};
    \node[packet, target] (mac3) at (4.5,-3.25) {$MAC_3$};
    \foreach \i in {1,2,3} {
        \node[above right] at (mac\i.north west) {\scriptsize EA};
    }

    \draw[latex-latex, adversary, thick, bend left=45]
        (mac1.east)
        to node[right] {\faIcon{mobile-alt} ?}
        (mac3.east);

    \coordinate (min_hr) at (3.25,.5);
    \coordinate (max_vt) at (3.5,.5);
    \coordinate (min_vt) at (3.5,-3.75);
    \coordinate (max_hr) at (6.4,.5);
    
    \begin{pgfonlayer}{bg}
        \fill[adversary, opacity=0.2]
          ($ (min_hr |- adversary.north) + (0,.17) $) [rounded corners=5pt] --
          ($ (max_hr |- adversary.north) + (0,.17) $) [sharp corners] --
          (max_hr |- adversary.south) [sharp corners] --
          (min_hr |- adversary.south) [rounded corners=5pt] -- cycle;
          
        \draw[
            adversary,
            ultra thick, rounded corners=5pt,
        ] (min_hr |- min_vt) rectangle (max_hr |- max_vt);
        \draw[adversary, ultra thick] (min_hr |- adversary.south) -- (max_hr |- adversary.south);
    \end{pgfonlayer}
    
\end{tikzpicture}
    \caption{Threat model scenario.}
    \label{fig:threat-model}
\end{figure}

We consider a passive adversary located within radio range of the target device and capable of capturing BLE advertising packets transmitted on the advertising channels. The adversary does not interact with devices, does not perform active probing, and does not possess any cryptographic material such as the Identity Resolution Key (IRK).

The target device employs Resolvable Private Addresses (RPAs), which rotate periodically. We denote the time-varying address of the target device as ${RPA}_{TD}(t)$. In addition to the target, the environment contains multiple unrelated devices emitting advertising traffic using their own addresses, which may or may not rotate, denoted as ${A}_{NTD}(t)$.

As illustrated in Figure~\ref{fig:threat-model}, the adversary observes a stream of advertising packets associated with multiple MAC addresses (e.g., ${MAC}_1$, ${MAC}_2$, ${MAC}_3$). Due to address rotation, these addresses cannot be directly linked to a stable device identity. The adversary's objective is therefore to determine whether observed advertising packets belong to a specific target device despite MAC address randomisation.

We model the attack as a two-phase process. During an initial observation period, i.e., a \emph{training phase}, the adversary collects advertising packets known to be emitted by the target device, as well as advertising packets from unrelated nearby devices.\footnote{{Such knowledge arises, for instance, because the adversary had previous physical access to the device, temporarily observed it in isolation, or was once a trusted party before losing its pairing privileges.}} These packets form positive and negative samples for training a model that captures device-specific characteristics observable at the advertising layer. In a \emph{deployment phase}, the adversary applies the trained model to newly observed traffic to determine whether the target device is present under previously unseen RPAs.

For the purpose of dataset construction only, we assume the availability of a trusted device that has been paired with the target and possesses the corresponding IRK. This device can resolve RPAs to a stable identity and is used exclusively to generate ground-truth labels. This capability is not available to the adversary during the attack.
\section{Related Work}
\label{sec:related-work}

As BLE became widely adopted in IoT and consumer devices, Bluetooth security and privacy research increasingly focused on whether its privacy mechanisms prevent long-term tracking.
Several studies have shown that address randomisation is not implemented consistently across devices, and that implementation flaws may expose permanent addresses~\cite{b2}. Infrequent address rotation can also enable tracking through temporal and signal-strength continuity~\cite{b3}. Even under standard-compliant behaviour, successive addresses can be linked using weak identifiers such as advertising intervals and limited simultaneous rotations~\cite{b4}.





\subsection{Signature-based Device Re-identification}
\label{sec:signature-based-identification}

Several works have shown that BLE advertising traffic can be exploited for device re-identification despite MAC address randomisation. Existing approaches can be broadly categorised into payload-based linkage, temporal and signal-based matching, and higher-layer fingerprinting.

Becker et al.~\cite{b13} show that many devices embed persistent identifying tokens within advertising payloads. They propose an address-carryover algorithm that exploits asynchronous changes between payload fields and randomised addresses to associate successive MAC addresses under continuous observation. Complementing this line of work, Akiyama et al.~\cite{b9,b18} propose identification methods based on temporal and signal-strength characteristics of advertising packets. Their approaches formulate MAC association as a matching problem, leveraging reception time and RSSI patterns to link addresses even when device-specific payload tokens are unavailable.

Boussad et al.~\cite{b10} introduce McMatcher, which constructs symbolic representations of RSSI time series to match random MAC addresses without supervised model training. Similarly, Despres et al.~\cite{b16} show that transmission characteristics, such as signal strength and advertising intervals, can be used to link rotating MAC addresses for the purpose of detecting malicious BLE trackers.


At a higher protocol layer, Celosia et al.~\cite{b15} show that GATT profiles can serve as fingerprints: services, characteristics, and associated metadata often remain sufficiently stable to uniquely identify devices, further limiting BLE privacy mechanisms.

Collectively, these studies show that BLE advertising-layer metadata, payload structure, and signal-level features can provide sufficient continuity to associate randomised BLE MAC addresses, even when address rotation is correctly implemented. However, these approaches typically rely on manually designed signatures, specialised matching procedures, or continuous observation of the target device.

\subsection{Machine Learning in Wireless Security}
\label{sec:machine-learning-in-ble-security}

Machine learning has been applied to device fingerprinting and identification problems in several wireless domains. For instance, Sobot et al.~\cite{sobot2022machine} investigate supervised learning techniques for device identification using wireless fingerprints extracted from Wi-Fi and narrowband IoT communications. Similarly, Bezawada et al.~\cite{bezawada2018iotsense} propose IoTSense, a behavioural fingerprinting framework that uses machine learning models trained on network-traffic features to identify IoT device types. These works show that learning-based methods can distinguish devices from observable communication features, but they target different wireless technologies or device-type identification rather than BLE address re-identification.

In the context of BLE, machine learning has primarily been applied to detect anomalous behaviour or attacks within BLE communication, rather than to link anonymised identifiers to a persistent device identity. Lahmadi et al.~\cite{b11} propose reconstruction and classification techniques to detect Man-in-the-Middle (MitM) attacks in BLE networks. Existing surveys of BLE security~\cite{b1,b14,b19} highlight growing interest in data-driven detection mechanisms. Address re-identification is not considered among the supervised learning problems.

\subsection{Advertising Behaviour Analysis}
\label{sec:Adv-behaviour-analysis}
{The analysis of device advertising behaviour is a critical metric for evaluating the LE privacy feature. Celosia et al. \cite{Celosia2019SavingPrivateAddresses} establish a systematic evaluation framework, identifying several device implementation flaws, such as address lifetime non-compliance and the non-uniform distribution of random static addresses, which undermine BLE privacy provisions. Complementing their work in \cite{Celosia2020DiscontinuedPrivacy}, the authors provide a deeper analysis of Apple’s ecosystem proximity-based services. The analysis proved that, despite the use of address randomisation, advertising traffic can leak personal and behavioural user actions.}

Overall, prior work demonstrates that BLE address randomisation can be circumvented using payload inspection, temporal analysis, or signal-level features. However, these approaches typically rely on handcrafted heuristics, continuous observation of the target device, or specialised matching procedures.

In contrast, our work formulates BLE device re-identification as a supervised learning problem over advertising-layer features and evaluates whether standard machine learning models can learn target-specific patterns without manual signature design. Additionally, we provide a practical methodology to construct labelled BLE traces, enabling systematic evaluation of learning-based attacks.

\section{BLE Trace Collection}
\label{sec:data-collection}

In this section, we describe the experimental setup and data collection methodology used to construct our labelled BLE advertising trace dataset.

\subsection{Testbed and Data Collection}
\label{sec:testbed}

Our testbed consists of three entities:
\begin{itemize}
    \item a \emph{target device}, which is the subject of our tracking;
    \item a \emph{trusted device}, which has been previously paired with the target device and is thus able to resolve the target device's RPA through the corresponding IRK;
    \item an \emph{adversary device} which is not paired with the target device and therefore can only observe the broadcast RPAs without resolving them.
\end{itemize}
The trusted device is used only for dataset construction: its resolution capability is used exclusively to generate ground-truth labels and is not available to the adversary during the attack.

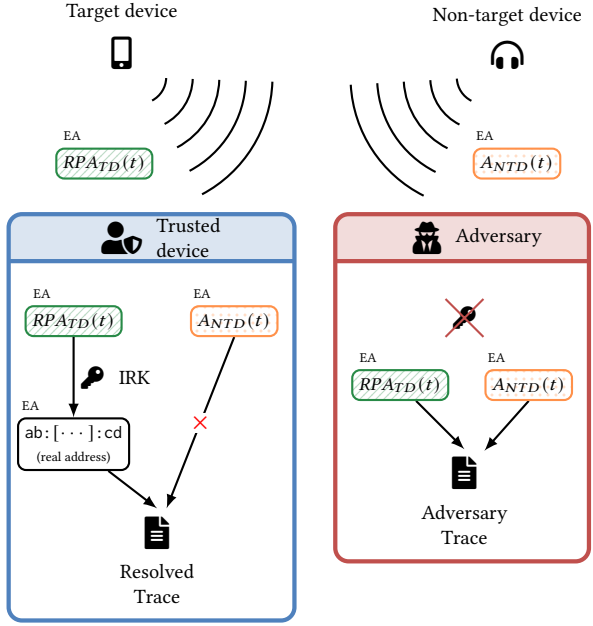
\begin{figure}
    \centering
    \pgfdeclarelayer{bg}
\pgfsetlayers{bg,main}

\definecolor{trusted}{HTML}{4F81BD}
\definecolor{adversary}{HTML}{C0504D}
\definecolor{target}{HTML}{288A47}
\definecolor{target-bg}{HTML}{BEDBC7}
\definecolor{nontarget}{HTML}{FF914A}
\definecolor{nontarget-bg}{HTML}{FFDEC8}

\tikzset{
  packet/.style={
    draw,
    thick,
    font=\small,
    rounded corners=3pt,
    align=left
  },
  target/.style = {
    draw = target,
    pattern = north east lines,
    pattern color = target-bg
  },
  nontarget/.style = {
    draw = nontarget,
    pattern = dots,
    pattern color = nontarget-bg
  }
}

\begin{tikzpicture}[scale=0.85, transform shape]

    \draw
        (0,0) node[
                label={[align=left]east:{Trusted \\ device}}
            ] (trusted) {\huge \faIcon{user-shield}}
        ++(-0.75,-1.3) node[packet, target, label={above left:{\scriptsize EA}}] (trusted_td_addr) {$RPA_{TD}(t)$};
        
    \draw[-latex, thick]
        (trusted_td_addr)
        -- node[right, label={east:IRK}] (irk) {\large \faIcon{key}}
        ++(0,-1.5) node[below, packet, label={above left:{\scriptsize EA}}, align=center] (trusted_real_addr) {\texttt{ab:}$[\cdots]$\texttt{:cd} \\ \scriptsize (real address)};

    \draw[-latex, thick]
        (trusted_real_addr)
        --
        ++(1.3,-1) node[below, label={[align=center]south:{Resolved \\ Trace}}] (resolved_trace) {\huge \faIcon{file-alt}};

    \draw[-latex, thick]
        (trusted)
        ++(1.75,-1.3) node[packet, nontarget, label={above left:{\scriptsize EA}}] (trusted_ntd_addr) {$A_{NTD}(t)$}
        (trusted_ntd_addr.south)
        -- node[red, fill=white, inner sep = 0.05cm] {\Large $\times$}
        (resolved_trace);
        
    \coordinate (min_hr) at ($ (trusted_td_addr.west) + (-0.25, 0) $);
    \coordinate (max_vt) at (trusted.north);
    \coordinate (min_vt) at ($ (resolved_trace.south) + (0,-1) $);
    \coordinate (max_hr) at ($ (trusted_ntd_addr.east) + (0.25,0) $);
    \begin{pgfonlayer}{bg}
        \fill[trusted, opacity=0.2]
          (min_hr |- trusted.north) [rounded corners=5pt] --
          (max_hr |- trusted.north) [sharp corners] --
          (max_hr |- trusted.south) [sharp corners] --
          (min_hr |- trusted.south) [rounded corners=5pt] -- cycle;
          
        \draw[
            trusted,
            ultra thick, rounded corners=5pt,
        ] (min_hr |- min_vt) rectangle (max_hr |- max_vt);
        \draw[trusted, ultra thick] (min_hr |- trusted.south) -- (max_hr |- trusted.south);
    \end{pgfonlayer}

    \node[label={east:{Adversary}}] at (4.75,0) (adversary) {\huge \faIcon{user-secret}};

    \draw[-latex, thick]
        (adversary.east)
        ++(0.25,-1.25) node (no_irk) {\large \faIcon{key}}
        (no_irk.south)
        ++(-1,-.75) node[packet, target, label={above left:{\scriptsize EA}}] (adversary_td_addr) {$RPA_{TD}(t)$}
        (adversary_td_addr)
        --
        ++(1,-1) node[below, label={[align=center]south:{Adversary \\ Trace}}] (adversary_trace) {\huge \faIcon{file-alt}};

    \draw[-latex, thick]
        (no_irk.south)
        ++(1,-.75) node[packet, nontarget, label={above left:{\scriptsize EA}}] (adversary_ntd_addr) {$A_{NTD}(t)$}
        (adversary_ntd_addr.south)
        -- (adversary_trace);

    \draw[adversary, thick] (no_irk.south east) -- (no_irk.north west) (no_irk.south west) -- (no_irk.north east); 

    \coordinate (min_hr) at ($ (adversary_td_addr.west) + (-0.25, 0) $);
    \coordinate (max_vt) at (adversary.north);
    \coordinate (min_vt) at ($ (adversary_trace.south) + (0,-1) $);
    \coordinate (max_hr) at ($ (adversary_ntd_addr.east) + (0.25,0) $);
    \begin{pgfonlayer}{bg}
        \fill[adversary, opacity=0.2]
          (min_hr |- adversary.north) [rounded corners=5pt] --
          (max_hr |- adversary.north) [sharp corners] --
          (max_hr |- adversary.south) [sharp corners] --
          (min_hr |- adversary.south) [rounded corners=5pt] -- cycle; 
          
        \draw[
            adversary,
            ultra thick, rounded corners=5pt,
        ] (min_hr |- min_vt) rectangle (max_hr |- max_vt);
        \draw[adversary, ultra thick] (min_hr |- adversary.south) -- (max_hr |- adversary.south);
    \end{pgfonlayer}

    \draw[decoration={expanding waves,angle=30},decorate, thick]
        (0,2.5) node[above, label={north:Target device}] (target) {\huge \faIcon{mobile-alt}}
        (target.south east)
        -- node[packet, target, below left=.5cm and .75cm, label={above left:{\scriptsize EA}}] {$RPA_{TD}(t)$}
        (2.2,1.3);

    \draw[decoration={expanding waves,angle=30},decorate, thick]
        (6,2.5) node[above, label={north:Non-target device}] (nontarget) {\huge \faIcon{headphones-alt}}
        (nontarget.south west)
        -- node[packet, nontarget, below right=.5cm and .75cm, label={above left:{\scriptsize EA}}] {$A_{NTD}(t)$}
        (3.8,1.3);
    
\end{tikzpicture}
    \caption{Testbed configuration for data collection.}
    \label{fig:testbed}
\end{figure}

As described in Section~\ref{sec:threat-model}, the target device emits EA packets using a time-varying address (i.e., $RPA_{TD}(t)$), while unrelated (non-target) background devices emit advertising traffic using their own addresses ($A_{NTD}(t)$), which may or may not rotate.

We configure the testbed such that both the trusted device and the adversary are within radio range and can observe the EA packets broadcast by the target device,  as illustrated in Figure~\ref{fig:testbed}.

For each target device, we simultaneously collect two BLE traces: one \emph{adversary trace} captured by the adversary, and one \emph{resolved trace} recorded by the trusted device. The trace temporal alignment ensures that packets observed in the adversary trace can be matched to the corresponding ground-truth information derived from the resolved trace.\footnote{Some packets may be missing due to channel loss or capture limitations. This may introduce asymmetry between traces, but our matching procedure is robust to partial observations.}

Since the trusted device is the sole possessor of the IRK, only it can resolve all $RPA_{TD}(t)$ to the underlying stable identity.
We do not filter the adversary trace, which contains both target and non-target traffic. In contrast, the resolved trace collected by the trusted device is restricted to packets emitted by the target device and is used solely for ground-truth construction.

\subsection{Ground Truth Construction}
\label{sec:trace-labelling}

The key idea is to use the resolved trace to extract Advertising Data (AD) structures specific to the target device, and then identify the corresponding packets in the adversary trace.

Given the adversary and resolved traces, we construct ground-truth labels for the adversary trace such that EA packets emitted by the target device are distinguished from those generated by background devices, thereby forming the basis for subsequent supervised learning.
To accomplish this, we design an algorithm to label each packet in the adversary traces using the corresponding resolved traces.


The algorithm unfolds as follows:
\begin{enumerate}
    \item From the resolved trace, for each EA packet emitted by the target device, we extract the AD and construct a reference set $\mathcal{R}_{AD}$ of AD structures.\footnote{We discard EA messages with an empty AD field.}
    \item We group the packets in the adversary trace by advertised MAC address. For each address $m$, we compute the total number of packets $n_{total}(m)$ containing any AD and the number of packets whose AD is contained in $\mathcal{R}_{AD}$,  denoted as $n_{match}(m)$.
    We then define the AD appearance ratio as
    $ \rho(m) = \frac{n_{\text{match}}(m)}{n_{\text{total}}(m)}.$
    $\rho(m)$ represents the fraction of packets sent by $m$ whose AD field exactly matches one of the AD structures observed in the resolved trace.
    \item If $\rho(m)\geq\theta$, where $\theta$ is a fixed threshold, we consider $m$ as belonging to the target device. This process yields a reference set $\mathcal{R}_{RPA}$ of RPAs belonging to the target device.
    \item We label EA packets whose address is in $\mathcal{R}_{RPA}$ as \emph{target}, while others are labelled as \emph{non-target}.
\end{enumerate}

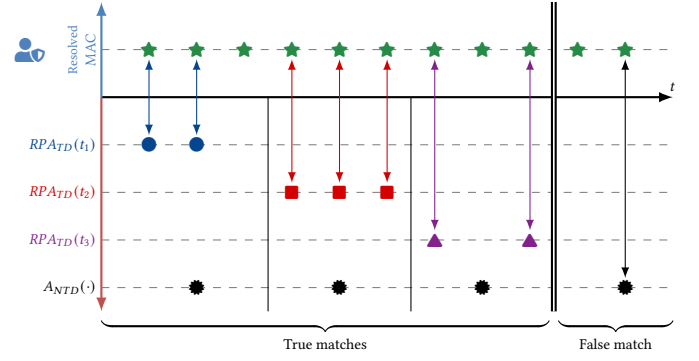
\begin{figure}
    \centering
    \definecolor{trusted}{HTML}{4F81BD}
\definecolor{adversary}{HTML}{C0504D}
\definecolor{real}{HTML}{288A47}
\definecolor{anon1}{HTML}{03468F}
\definecolor{anon2}{HTML}{D20000}
\definecolor{anon3}{HTML}{82218b}

\begin{tikzpicture}[scale=0.63, transform shape]

    \draw[-latex, thick] (0,0) -- (12,0) node[above] {$t$};

    \draw[trusted, -latex, thick] (0,0) -- node[midway, above, rotate=90, align=center] {\small Resolved \\ MAC} (0,2);
    \node[trusted] at (-1.5,1) {\huge \faIcon{user-shield}};
    \draw[gray, dashed] (0,1) -- (12,1);
    
    \foreach \x in {0,...,10}
        \node[real] at (1+\x,1) {\faIcon{star}};

    \draw[adversary, -latex, thick] (0,0) -- (0,-4.5);
    
    \foreach \i in {1,2,3}
        \draw[gray, dashed] (0,-\i) node[left, anon\i] {$RPA_{TD}(t_\i)$} -- (12,-\i);
    \draw[gray, dashed] (0,-4) node[black, left] {$A_{NTD}(\cdot)$} -- (12,-4);

    \foreach \x in {1,2} {
        \node[anon1] at (\x,-1) {\faIcon{circle}};
        \draw[anon1, latex-latex] (\x,.8) -- (\x,-.8);
    }
    
    \foreach \x in {4,5,6} {
        \node[anon2] at (\x,-2) {\faIcon{square}};
        \draw[anon2, latex-latex] (\x,.8) -- (\x,-1.8);
    }
    
    \foreach \x in {7,9} {
        \node[anon3, rotate=90] at (\x,-3) {\faIcon{play}};
        \draw[anon3, latex-latex] (\x,.8) -- (\x,-2.8);
    }

    \foreach \x in {2,5,8,11} 
        \node at (\x,-4) {\faIcon{certificate}};
    \draw[latex-latex] (11,.8) -- (11,-3.8);

    \foreach \x in {3.5,6.5}
        \draw (\x,0) -- (\x,-4.5);
    \draw[double, thick] (9.5,2) -- (9.5,-4.5);

    \draw[decorate, thick, decoration={calligraphic brace, amplitude=4pt, mirror}] (0,-4.7) -- node[midway, below = .25cm] {True matches} (9.4,-4.7);

    \draw[decorate, thick, decoration={calligraphic brace, amplitude=4pt, mirror}] (9.6,-4.7) -- node[midway, below = .25cm] {False match} (12,-4.7);
\end{tikzpicture}
    \caption{GT matching algorithm. For each possible MAC (bottom), we count the number of EAs that match in the resolved trace (top). We discard false matches (device on the bottom) that have low match ratio.}
    \label{fig:gt-matching}
\end{figure}

The threshold $\theta$ mitigates false associations caused by background devices that occasionally emit advertising data identical to that of the target device. This situation may occur when short or generic AD structures are shared across devices, as illustrated in Figure~\ref{fig:gt-matching}. Packet loss or capture limitations may also cause asymmetries between the adversary and resolved traces.

To select $\theta$, we analyse the distribution of $\rho(m)$ for all addresses exhibiting at least one match with $\mathcal{R}_{AD}$. We observe that all target-device addresses have $\rho(m) > 0.5$, while the only spurious matches are associated with addresses for which a single advertisement matches the resolved trace. We therefore set $\theta=0.5$, which preserves all target RPAs while removing these false associations.


\subsection{Dataset Overview}
\label{sec:dataset}

\begin{table}
    \centering
    \caption{Device breakdown by type}
    \begin{tabular}{lc}
        \toprule
        \textbf{Target type} & \textbf{\# Devices} \\
        \midrule        
        Earbuds / Headphones & 7 \\
        Phone & 2 \\
        Smartwatch & 1 \\
        Tracker tag & 1 \\
        Laptop & 1 \\
        Tablet & 1 \\
        Development board & 1 \\
        \midrule
        \textbf{Total} & \textbf{14} \\
        \bottomrule
    \end{tabular}
    \label{tab:device-types}
\end{table}

We collected BLE traces in different environments, each containing one target device and several nearby BLE devices that were not controlled, and in some cases not even known, by the experimenter. We first collected three traces in our lab, and then asked Master's students to replicate the same data collection procedure in their own environments. Eleven students completed the task successfully, resulting in a final dataset of 14 traces, each centred on a unique target device.

\subsubsection{{Trace analysis}}
The target devices include smartphones, smartwatches, laptops, tablets, earbuds, tracker tags, and development boards. Table~\ref{tab:device-types} summarises their distribution by category.
%
%
Each trace lasts at least 30 minutes, which is sufficient to observe multiple address rotations per device under typical RPA refresh intervals of 15 minutes.

\begin{figure}
    \centering
    \includegraphics[width=.85\columnwidth]{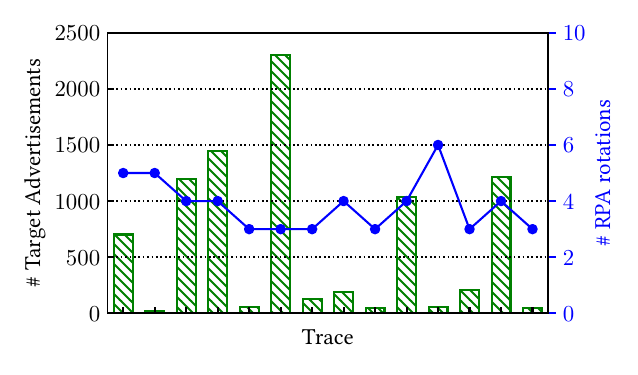}
    \caption{Trace composition before noise injection. The bars show the number of target (positive) BLE advertisements in each trace, while the line reports the number of distinct RPAs observed for the target device.}
    \label{fig:target-advertisements}
\end{figure}

Using the labelling strategy described in Section~\ref{sec:trace-labelling}, we identify the packets emitted by the target device and use them as positive samples in the final evaluation dataset.
Figure~\ref{fig:target-advertisements} summarises the contents of each trace. For each target device, we report the number of advertisements emitted by the target device (green bar), while the blue line reports the number of distinct target RPAs observed during the capture period.
As expected, the number of advertisements sent by each target device varies substantially across traces, because it depends on the density and advertising behaviour of each device. Each trace contains at least three RPAs for the target device, enabling evaluation across multiple address rotations.

All remaining advertising packets, i.e., packets not labelled as belonging to the current target device, are then collected into a \emph{noise bank}. The noise bank therefore contains real background traffic observed across the collected traces, including non-target devices that were present during the target device capture.

For each observed RPA epoch of the target device, we inject advertisements from a varying number $N$ of unrelated devices and temporally align them with the corresponding address-rotation interval. $N$ allows us to control the amount of background traffic seen by the classifier while preserving realistic advertising behaviour: all negative samples originate from real BLE devices and retain their original packet contents and transmission patterns.
{We detail the device sampling strategy in Section~\ref{sec:validation-strategy}.}

\subsubsection{{Noise bank characterisation}}
We construct the noise bank from advertising traffic emitted by both target and non-target devices observed across the collected traces. To avoid including sparse advertisers, we only include devices that broadcast at least 10 advertising packets during their observation period.

\begin{figure}
    \centering
    \begin{subfigure}{\columnwidth}
        \centering
        \includegraphics[width=.85\columnwidth]{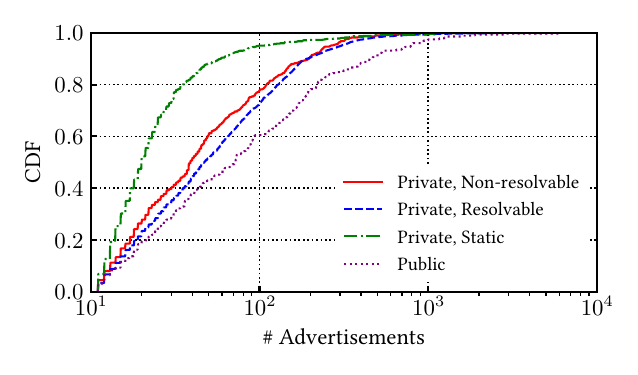}
        \caption{Number of advertisements}
        \label{fig:noise-bank-size}
    \end{subfigure}
    \begin{subfigure}{\columnwidth}
        \centering
        \includegraphics[width=.85\columnwidth]{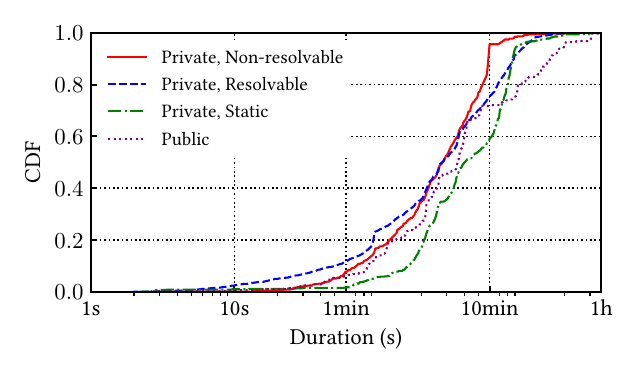}
        \caption{Observation duration}
        \label{fig:noise-bank-duration}
    \end{subfigure}
    \caption{Per-device characterisation of the noise bank, grouped by BLE address type.}
    \label{fig:noise-bank-characterisation}
\end{figure}

In total, the noise bank contains 345,283 advertising packets broadcast by 3,884 distinct BLE addresses.
Of these, 56.4\% of them are resolvable private, 19.6\% are static private, 16.9\% are non-resolvable private and 7.1\% of them are of public type.
These statistics show that, while most traffic is associated with RPAs, there is still a substantial fraction of traffic associated with static, public or non-resolvable addresses.\footnote{These statistics are computed over BLE address instances rather than physical devices. Since the same RPA-enabled device may contribute multiple addresses over time, these fractions are not indicative of device-level adoption rates.}
Nevertheless, their inclusion is important for emulating realistic BLE background traffic, since a passive observer is exposed to a mixture of address behaviours rather than only to devices implementing address rotation.

Figure~\ref{fig:noise-bank-characterisation} provides further characterisation of the background devices in the noise bank by separating them according to address type.
Figure~\ref{fig:noise-bank-size} shows the distribution of the number of packets emitted by each device, highlighting the biassed nature of BLE traffic. While the average address contributes approximately 90 packets, we observe a small amount of addresses that emit more than 1,000 packets during trace collection. Public address types contribute substantially more advertisements compared to the other address types, likely because they do not rotate over time. However, private static addresses, which are expected to exhibit the same behaviour, contribute way fewer packets.

Figure~\ref{fig:noise-bank-duration} shows the broadcast duration distribution, computed as the time passed between the first and last packet sent by the given BLE address.
It shows that most devices are visible for less than 10 minutes.
This behaviour is expected in real BLE environments, where devices may enter or leave radio range, advertise intermittently, or remain continuously present during the capture.
As also expected, public and private static addresses are present for longer times on average compared to their (non-)resolvable counterparts, since the latter are periodically rotated and thus split the same physical device's activity across multiple addresses.

\section{Classification Framework}
\label{sec:classification}

Starting from the labelled traces, we formulate device linkage as a supervised binary classification problem over individual BLE advertising packets. Each packet observed in the passive trace is represented by a feature vector and assigned a label according to the ground-truth. The classifier predicts whether a given packet was emitted by the target device (\textit{target}) or by any other device in the environment (\textit{non-target}).

The ultimate objective of the adversary is not to classify individual packets, but to determine whether the target device is present in the observed environment despite MAC address randomisation. Since the adversary does not know which RPAs correspond to the target device, this requires identifying which observed MAC addresses could plausibly belong to it.

To this end, we adopt a two-stage approach. First, the classifier identifies packets likely emitted by the target device. Second, packet-level predictions are aggregated at the MAC-address level: for each observed address, we consider all associated packets and determine whether the address could correspond to an instance of $\mathrm{RPA}_{TD}(t)$. Thus, packet-level classification provides the evidence used to infer device presence and identify candidate target addresses.

In the following, we describe the feature extraction process and justify the choice of the classifier used to evaluate the feasibility of packet-level discrimination under MAC randomisation.

\subsection{Feature Extraction}
\label{sec:feature-extraction}

For each advertising packet, we extract a set of features that are directly observable by a passive adversary and that may exhibit persistence across RPA rotations.  

For all packets, we extract protocol fields and capture metadata associated with each advertising event. These include header-level attributes exposed by the BLE stack, such as event type, connectable and scannable flags, payload length, and received signal strength when available. Although these fields do not uniquely identify a device, prior work suggests that they may exhibit consistent patterns across address changes and therefore contribute to device re-identification.

For packets containing non-null advertising payloads, we extract the Advertising Data (AD) field and encode its content using overlapping byte-level 3-grams derived from the hexadecimal byte sequence. This byte-level representation avoids assuming that the adversary is a Bluetooth expert or can parse the semantics of AD structures, which may contain both standard fields and manufacturer-specific data. By modelling stable fragments of the payload rather than relying on specific field matches, the classifier can autonomously learn structural regularities that persist across packets and RPA epochs.

All extracted features are encoded into a sparse vector representation suitable for supervised learning. The feature space combines metadata attributes and payload-derived n-grams. Feature extraction is strictly limited to attacker-observable information: no information derived from the resolved trace, address resolution, or pairing state is included as an input feature.

Given the class imbalance of our dataset, we train and evaluate the classifiers under skewed class distributions.

\subsection{Classifier Choice}
\label{sec:classifier-choice}

We employ decision tree classifiers as a proof-of-feasibility approach to distinguish between \textit{target} and \textit{non-target} devices.

While simple, decision trees are well-suited to this setting. They support heterogeneous feature types, including continuous or discretised measurements (e.g., RSSI), binary indicators (e.g., advertising flags), and high-dimensional sparse features derived from payload n-grams. Additionally, they can model non-linear interactions between metadata and payload features, which is important when device-specific behaviour arises from combinations of advertising-layer attributes rather than single invariant fields. Finally, they provide interpretability through explicit splitting rules and feature importance scores, enabling inspection of which observable features drive re-identification.

Overall, our goal is not to optimise classification performance through complex model architectures, but to assess whether standard supervised learning techniques are sufficient to automate an attack that previously required manual signature construction.

We evaluate this framework using a group-aware validation strategy that holds out entire RPA epochs, as described in Section~\ref{sec:validation-strategy}.

For comparison, we also evaluate a simple exact-payload matching strategy as a baseline. It classifies a packet as belonging to the target only if its complete AD payload exactly matches a payload previously observed for the target.

\section{Evaluation}
\label{sec:evaluation}

In this section, we evaluate the effectiveness of the proposed classification framework.

\subsection{Validation Strategy}
\label{sec:validation-strategy}


To evaluate whether the proposed classification framework generalises across address rotations, we adopt a nested, group-aware validation strategy based on the target device's RPA epochs. For a target device observed across $K$ RPA epochs, each validation round trains the classifier on {2} epochs (one used for training, one used for validation), assuming that the attacker has limited information about the target device. We evaluate the trained model on the remaining {$K-2$} epochs (used for testing). This guarantees that the held-out epochs correspond to RPAs that were never seen during training/validation, thereby testing whether the learned advertising-layer features persist across address rotations.

The process is performed independently for each target device: for each target, we train a separate binary classifier to distinguish that device from background traffic. The objective is therefore target-device re-identification, not cross-device classification.

\begin{table}
    \centering
    \caption{Hyperparameter Selection}
    \begin{tabular}{rl}
         \toprule
         \textbf{Parameter} & \textbf{Values tested} \\
         \midrule
         \texttt{max\_depth} & 4, 6, 8, 10, None \\
         \texttt{min\_samples\_leaf} & 1, 5, 10 \\
         \texttt{min\_samples\_split} & 2, 4, 8 \\
         \bottomrule
    \end{tabular}
    \label{tab:hyperparameters}
\end{table}

The procedure unfolds as follows:
\begin{enumerate}
    \item \label{itm:rpa-group}
    \textbf{Group by RPA epoch}: For each target device, we partition the labelled scenario into $K$ groups, where each group corresponds to one RPA epoch of the target device. A group contains all packets emitted by the target device under that RPA.

    \item \label{itm:fixed-test-groups}
    {\textbf{Construct fixed test groups}:} For each group, we construct a \textit{fixed} test group by injecting traffic from $N$ unrelated BLE addresses sampled from the noise bank. These groups are reused across training settings. We fix $N=50$, ensuring that all models are tested under identical, maximally challenging conditions.
    
    \item \label{itm:outer-split}
    {\textbf{Outer validation loop}}: We apply a leave-$P$-groups-out strategy, with $P=K-2$, thereby training on two RPA epochs and testing on the remaining $K-2$. This represents a challenging scenario in which the attacker has labelled traffic from only two RPA rotations of the target device. 

    \item \label{itm:train-test-construction}
    {\textbf{Train-test scenario construction}:} For each split, we construct the training and testing sets separately. For the testing set, we extract the fixed test groups belonging to the held-out test epochs. This step ensures that, for a given split, all training noise settings are evaluated on the same noisy test groups. For the training set, we take the two selected epochs from the labelled trace before noise injection and inject traffic from $N\in\{10,20,30,40,50\}$ unrelated background addresses, allowing us to evaluate multiple noise conditions.\footnote{When injecting training noise, we exclude background addresses that appear in the corresponding test groups, thus preventing possible shortcuts caused by the same background address appearing in both training and testing.}
    The exact-payload baseline is evaluated using the same split: it stores the complete AD payloads observed for the target in the two training epochs and classifies a test packet as \textit{target} only if its complete AD payload matches one of them.
    
    \item \label{itm:hyperparam-selection}
    \textbf{Inner validation and hyperparameter selection}: Within each training split, we perform a grid search over the parameters listed in Table~\ref{tab:hyperparameters}. We use an inner group-aware cross-validation over the {2} training epochs, so that hyperparameters are selected without using packets from the held-out test epochs.
    
    \item \label{itm:training-eval}
    \textbf{Training and evaluation}: Using the selected hyperparameters, we train the final decision tree classifier on all training groups and evaluate its performance on the held-out test groups.
\end{enumerate}

This process yields {$\binom{K}{2}$} decision trees for each target device and noise setting. 
Each model classifies whether a given EA packet belongs to the target device based on its EA information.

\subsection{Packet-Level Classification Performance}
\label{sec:packet-level-classification}


We first evaluate the classifier's ability to distinguish individual advertisements emitted by the target device from those sent by background devices. Since the validation procedure holds out entire RPA epochs, successful classification indicates that the classifier has learned advertising-layer features that remain stable across address rotations. 

At the packet level, the classifier achieves a mean precision of 91.4\%, a mean recall of 98.6\%, and a mean F1-score of 92.6\%.
The high recall shows that the classifier correctly retrieves practically all advertisements of the target (true positives), even when the model is trained on only two RPA epochs.
In comparison, the exact-payload matching baseline achieves a packet-level recall of 52.9\%. The decision tree therefore recovers substantially more target advertisements from previously unseen RPA epochs, showing that complete AD payloads are not always stable across observations, while recurring payload fragments remain sufficiently persistent for the classifier to exploit them.
These results are particularly strong considering the conservative evaluation setting: each model is trained using only two RPA epochs of the target device and just 10 noise devices (negative devices). This increases both class imbalance and the chance of observing advertising patterns similar to those of the target.

The lower precision is largely explained by the evaluation protocol. During testing, the classifier is evaluated against advertisements from 50 unrelated devices, resulting in several thousand negative packets for every validation split. Consequently, even a very low false-positive rate translates into a noticeable reduction in precision.  
We next analyse whether increasing the amount of background traffic
available during training affects the classifier's tendency to produce false positives.

\subsection{Impact of Training Noise}
\label{sec:noise-impact}

We vary the amount of background traffic injected into the training epochs $N \in \{10,20,30,40,50\}$. We keep the test scenario fixed across all training-noise settings with $N=50$ unrelated devices.

\begin{figure}[b]
    \centering
    \includegraphics[width=\linewidth]{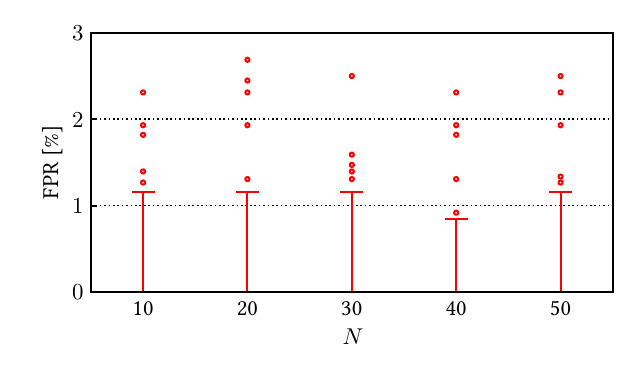}
    \caption{Distribution of split-level false positive rates (FPR) across training-noise settings.}
    \label{fig:fp-noise-distribution}
\end{figure}

\begin{figure*}
    \centering
    \includegraphics[width=\linewidth]{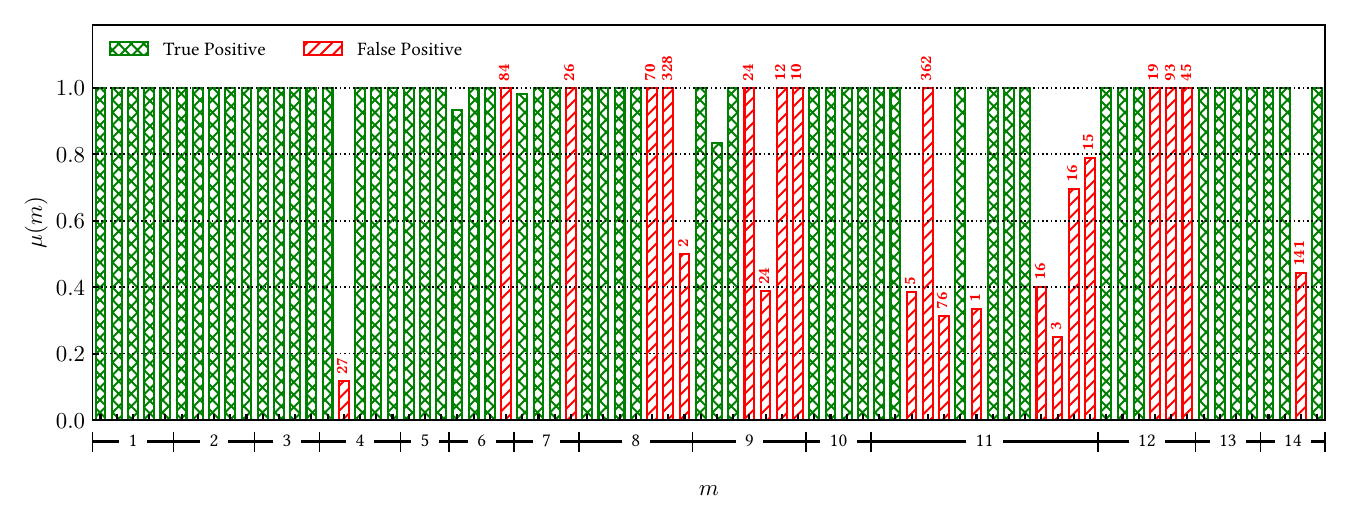}
    \caption{Values of $\mu(m)$ for MAC addresses $m$ whose packets have been classified as \textit{target} at least once when training with $N=10$ background devices. Green bars indicate the target device's true MAC addresses (true positives). Red bars indicate false positive MAC addresses. The numbers above each red bar indicate the number of positively classified packets. The numbers at the bottom indicate the trace number.}
    \label{fig:post-process-ratios}
\end{figure*}

Figure~\ref{fig:fp-noise-distribution} reports the distribution of the splits' false positive rates (FPR) for each training-noise setting $N$. We calculate the FPR as 
$
FPR = \frac{FP}{\lvert Noise\rvert},
$
or the ratio between the number of background advertisements incorrectly classified as target and the total number of background advertisements in the corresponding test split.
The boxplots include all outer validation splits. {Since more than 50\% of splits have no false positives, the median number of false positives is 0, and only the whiskers, which show the 95-percentile, are visible.}
Overall, the results show that the FPR remains low across all training-noise settings, with {95\%} of splits having an FPR less than {1.2\%}. Only in a few cases, the max FPR  grows to 2-3\% -- see the dots.

Interestingly, increasing the number of background devices used during training does not appear to produce a clear improvement in the false positive rate.
This suggests that the remaining false positives are not primarily caused by insufficient negative training samples. Instead, they appear to arise from specific advertisements whose advertising-layer features resemble those of the target device, which the single decision tree has a hard time separating.

Recall that $N$ controls the number of injected background devices' addresses, but not the absolute number of packets they generate.
As shown in Figure~\ref{fig:noise-bank-size}, each device contributes a highly variable number of advertisements, so two training splits with the same value of $N$ may still contain different amounts of negative traffic.

We also investigated whether the absolute number of injected background packets correlates with the FPR. However, we observe little relationship. This suggests that false positives are not primarily driven by the sheer quantity of negative packets available during training and that including additional packets does not necessarily add new discriminative information. Many advertisements emitted by the same background address share identical or highly similar advertising data, so increasing the number of packets may mostly repeat already observed patterns rather than expose the classifier to meaningfully new negative examples. More expressive ensemble models may further reduce these remaining false positives, which we leave for future work.

Given this strong packet-level performance, we next evaluate whether these predictions are sufficient to identify the target device at the MAC-address level.

\subsection{From Packet-Level Detection to Device Identification}
\label{sec:rpa-grouping}

The adversary's objective is to identify whether an observed MAC address corresponds to the target device. Since classification is performed at the packet level, predictions must be aggregated to infer address-level information.

We use a simple aggregation rule analogous to the one used for ground-truth construction. For each observed MAC address $m$, we compute the fraction of packets classified as \textit{target} 
$
\mu(m) = \frac{n_{\text{pred}}(m)}{n_{\text{total}}(m)},
$
where $n_{\text{pred}}(m)$ is the number of packets associated with $m$ classified as \textit{target}, and $n_{\text{total}}(m)$ is the total number of observed packets for that address.

We then mark $m$ as a candidate $\mathrm{RPA}_{TD}(t)$ if it satisfies two conditions:
    $n_{\text{total}}(m) \geq n_{\min}$, and
    $\mu(m) \geq \tau$,
where $n_{\min}$ filters out addresses supported by too few packets, and $\tau$ controls how consistently an address must be classified as target.

Figure~\ref{fig:post-process-ratios} reports the values of $\mu(m)$ for all addresses classified as \textit{target} at least once in the $N=10$ noise setting.
Green bars indicate \textit{true} target RPAs, while red bars indicate non-target RPAs that receive at least one positive packet-level prediction.
The figure shows that true target addresses generally exhibit high $\mu(m)$ values, typically above 80\%.
At the same time, several background addresses also receive positive predictions, some with $\mu(m)<80\%$.
Therefore, setting $\tau > 0.8$ correctly identifies all target addresses, and excludes 11 noise devices from being classified as positive, with just 11 false addresses identified as target out of a total of {2,904} negative devices, i.e., a 0.38\% of false device identification.\footnote{{Note that the total number of negative devices is not equal to 50 multiplied by the number of RPA epochs, as the same background address may appear in multiple target traces.}}
In comparison, under the same $\tau$ and $n_{min}$, the exact-AD payload-matching strategy achieves only a 35.2\% MAC-level recall, although lowering the false-positive rate to 0.037\%.\footnote{Lowering the baseline threshold to $\tau=0.5$, which maximises its MAC-level F1-score in our evaluation, increases its MAC-level recall to 46.3\%.}

Overall, this simple aggregation step shows that strong packet-level classification can translate into even stronger address-level re-identification. Even a modest number of correctly classified packets can be sufficient to flag candidate RPAs belonging to the target device, while the minimum-packet threshold limits false positives caused by sparse false-positive background addresses.
We leave a systematic exploration of MAC-level filtering criteria as future work.
\subsection{Feature Importance}
\label{sec:feature-importance}

A key advantage of decision trees is their interpretability. Unlike more complex models, decision trees provide explicit splitting rules and feature importance scores, allowing us to identify which observable attributes drive the classification process.

To analyse model behaviour, we extract feature-importance scores from the decision tree trained in each validation fold. We first average the scores across folds for each target device and normalise them so that the importances sum to one for that device. We then average the normalised per-device scores and normalise once more to obtain a global view of feature relevance across the dataset, giving each target device equal weight.

Our results reveal that classification is overwhelmingly driven by the content of the Advertising Data (AD) field: AD 3-grams account for 91.7\% of the total feature importance, indicating that the classifier primarily relies on recurring payload patterns to distinguish target-device packets. The single most influential 3-gram contributes 4.8\% of the overall importance, while the remaining importance is distributed across more than 119\,k other payload fragments.

In contrast, protocol-level metadata plays only a minor role in classification. The most relevant non-payload features are the AD length and RSSI, accounting for 3.6\% and 3.2\% of the total importance, respectively, followed by TX Power which accounts for 1.1\%. All other fields contribute 0.1\% or less, including the peer address type (0.1\%), secondary PHY (0.06\%), event type data flags (0.04\%), primary PHY (0.03\%) and advertising SID (0.02\%).

These results indicate that device re-identification is primarily driven by advertising-payload content rather than protocol metadata or signal-level characteristics. More importantly, the discriminative payload patterns are automatically discovered from byte-level n-grams, without requiring the adversary to parse AD structures or manually design device-specific signatures.

Overall, these results show that different devices expose different forms of persistent information at the advertising layer. In particular, specific values or patterns within the Advertising Data field can act as implicit identifiers, even in the absence of stable MAC addresses. This confirms findings of previous works: device re-identification is enabled by a combination of protocol behaviour and payload structure. Here, we show that both of them can be automatically learned by a supervised model.

\section{Discussion}
\label{sec:discussion}

Our results indicate that BLE device re-identification can be automated by learning persistent patterns in advertising-layer features.
However, the practical impact of this finding depends on the assumptions made about the attacker, the amount of labelled traffic available during training, and the diversity of the BLE environments considered in the evaluation. In this section, we therefore discuss the main limitations of our study before outlining possible countermeasures that could reduce the linkability of BLE advertising traffic.

\subsection{Limitations}
\label{sec:discussion}

While the results demonstrate the feasibility of automated BLE device re-identification, several limitations should be considered.

First, the threat model assumes a training phase in which the adversary observes the target device and obtains labelled positive and negative samples. The resulting model is therefore target-specific and does not generalise to arbitrary unseen devices. In practice, the attack requires prior exposure to the target device before it can be re-identified under new RPAs. Future work could relax this assumption by considering scenarios in which the adversary cannot isolate the target traffic during training, and instead trains classifiers for all devices observed during an initial collection period.

Second, classification is performed at packet level, while the adversary's objective is to identify devices at the MAC-address level. Packet-level errors may propagate during aggregation, potentially leading to missed target addresses or false candidate addresses. As shown in Section~\ref{sec:rpa-grouping}, this effect can be mitigated through thresholding, but a more systematic exploration of the $(n_{\min}, \tau)$ parameter space remains future work. Another direction is to train classifiers directly at the address or device level rather than aggregating packet-level predictions.

Third, our evaluation is based on 14 target devices and observation windows of at least 30 minutes per trace. Although the experiments cover different environments, device types, traffic conditions, and levels of background noise, further validation on larger datasets is required to assess robustness over longer time periods and across a broader range of devices.

Finally, our evaluation constructs controlled scenarios by injecting traffic from $N=50$ non-target devices sampled from the noise bank. This provides a reproducible way to evaluate the attack under realistic background traffic, but does not exhaust the full range of possible BLE environments, device densities, or mobility patterns.

\subsection{Possible Countermeasures}
\label{sec:countermeasures}

Previous work has shown that advertising payloads constitute a major source of linkability despite BLE MAC address randomization, as they often expose persistent information that can be used to re-identify devices across address changes~\cite{b13,b15}. Our results further reinforce this observation by showing that even simple machine learning models can automatically exploit such information without requiring protocol-specific knowledge or handcrafted signatures.

A promising direction is therefore to redesign the advertising process for privacy-sensitive devices. One possibility is to minimize the information exposed before connection establishment, postponing the disclosure of richer advertising data until after a trusted peer has authenticated or established a connection. Since paired devices can already resolve RPAs using the shared Identity Resolution Key (IRK), this approach would preserve normal operation while significantly reducing the information available to passive observers.

An alternative approach would be to cryptographically protect the advertising payload itself. Rather than transmitting advertising data in cleartext, devices could encrypt it using keys derived from the shared credentials established during pairing, allowing only authorized peers to recover the advertised information while preventing passive observers from exploiting persistent payload patterns. Such a solution would preserve the current discovery model for paired devices without requiring additional connection attempts.

Neither approach is suitable for all BLE applications. Devices whose purpose is public discovery, such as beacons, inherently trade privacy for discoverability. However, both appear well-suited to personal devices -- including smartphones, wearables, and peripherals -- whose primary objective is to prevent long-term tracking. Designing and evaluating such privacy-oriented advertising mechanisms, together with their impact on usability, interoperability, and energy consumption, remains an interesting future direction.
\section{Conclusions}
\label{sec:conclusion}

In this work, we investigated whether BLE device re-identification under MAC address randomisation can be automated using supervised machine learning. We formulated the problem as a packet-level classification task over advertising traffic and proposed a methodology to collect and label BLE traces using a dual-sniffer setup. Our evaluation on multiple commercial devices shows that even simple models can successfully learn device-specific patterns and identify target devices across address rotations in most cases.
In particular, the substantial performance gap with exact payload matching shows that relying on complete AD payload equality is unreliable in our traces. Although parts of the payload change across observations, recurring payload fragments remain sufficiently stable for the classifier to exploit.

Overall, our results confirm that BLE privacy protections remain limited. More than 15 years after the introduction of Resolvable Private Addresses, support across devices is still incomplete, and even when correctly implemented, other observable signals at the advertising layer enable reliable re-identification. This confirms that MAC address randomisation alone is insufficient to prevent tracking in practice.

\begin{acks}
This work has received funding from the Applied Sciences Italian Fund (Fondo Italiano per le Scienze Applicate---FISA) by the Italian Ministry of University and Research, under the AI4CTI project (grant agreement No. FISA-2023-00168).
\end{acks}

\bibliographystyle{ACM-Reference-Format}
\bibliography{bibliography}

@article{b1,
  author  = {C{\"a}sar, M. and Pawelke, T. and Steffan, J. and Terhorst, G.},
  title   = {A survey on Bluetooth Low Energy security and privacy},
  journal = {Computer Networks},
  volume  = {205},
  pages   = {108712},
  year    = {2022}
}

@misc{b2,
  author        = {Martin, J. and Mayberry, T. and Donahue, C. and Foppe, L. and Brown, L. and Riggins, C. and Rye, E. C. and Brown, D.},
  title         = {A study of MAC address randomization in mobile devices and when it fails},
  howpublished  = {arXiv preprint arXiv:1703.02874},
  year          = {2017}
}

@inproceedings{b3,
  author    = {Kalantar, G. and Mohammadi, A. and Sadrieh, S. N.},
  title     = {Analyzing the effect of Bluetooth Low Energy (BLE) with randomized MAC addresses in IoT applications},
  booktitle = {Proceedings of the IEEE International Conference on Internet of Things (iThings), GreenCom, CPSCom, SmartData},
  pages     = {27--34},
  year      = {2018}
}

@inproceedings{b4,
  author    = {Jouans, L. and Viana, A. C. and Achir, N. and Fladenmuller, A.},
  title     = {Associating the randomized Bluetooth MAC addresses of a device},
  booktitle = {Proceedings of the IEEE Consumer Communications and Networking Conference (CCNC)},
  pages     = {1--6},
  year      = {2021}
}

@article{b7,
  author  = {Zhang, Y. and Lin, Z.},
  title   = {Breaking BLE MAC address randomization with allowlist-based side channels and its countermeasure},
  journal = {ACM Transactions on Privacy and Security},
  year    = {2025}
}

@inproceedings{b8,
  author    = {Baccichet, G. and Innamorati, C. and Redondi, A. E. C. and Cesana, M.},
  title     = {MAC address de-randomization using multi-channel sniffers and two-stage clustering},
  booktitle = {Proceedings of the IEEE PIMRC},
  pages     = {1--6},
  year      = {2024}
}

@inproceedings{b9,
  author    = {Akiyama, S. and Taniguchi, Y.},
  title     = {Device identification in BLE packets from moving devices with randomized MAC addresses},
  booktitle = {Proceedings of the IEEE ICCE-Asia},
  pages     = {1--4},
  year      = {2023}
}

@inproceedings{b10,
  author    = {Boussad, Y. and Yang, Y. and Tomlinson, A. and Grant-Muller, S.},
  title     = {McMatcher: A symbolic representation for matching random BLE MAC addresses},
  booktitle = {Proceedings of the IEEE ICCE},
  pages     = {1--6},
  year      = {2024}
}

@inproceedings{b11,
  author    = {Lahmadi, A. and Duque, A. and Heraief, N. and Francq, J.},
  title     = {MitM attack detection in BLE networks using reconstruction and classification machine learning techniques},
  booktitle = {Joint European Conference on Machine Learning and Knowledge Discovery in Databases},
  pages     = {149--164},
  year      = {2020}
}

@article{b13,
  author  = {Becker, J. K. and Li, D. and Starobinski, D.},
  title   = {Tracking anonymized Bluetooth devices},
  journal = {Proceedings of Privacy Enhancing Technologies},
  year    = {2019}
}

@misc{b14,
  author       = {Wang, Z.},
  title        = {Securing Bluetooth Low Energy: A literature review},
  howpublished = {arXiv preprint arXiv:2404.16846},
  year         = {2024}
}

@inproceedings{b15,
  author    = {Celosia, G. and Cunche, M.},
  title     = {Fingerprinting Bluetooth-Low-Energy devices based on the generic attribute profile},
  booktitle = {Proceedings of the 2nd International ACM Workshop on Security and Privacy for the Internet-of-Things},
  pages     = {24--31},
  year      = {2019}
}

@inproceedings{b16,
  author    = {Despres, T. and Davis, N. and Dutta, P. and Wagner, D.},
  title     = {DeTagTive: Linking MACs to protect against malicious BLE trackers},
  booktitle = {Proceedings of the 2nd Workshop on Situating Network Infrastructure with People, Practices, and Beyond},
  pages     = {1--7},
  year      = {2023}
}

@inproceedings{b18,
  author    = {Akiyama, S. and Morimoto, R. and Taniguchi, Y.},
  title     = {A study on device identification from BLE advertising packets with randomized MAC addresses},
  booktitle = {Proceedings of the IEEE ICCE-Asia},
  pages     = {1--4},
  year      = {2021}
}

@inproceedings{b19,
  author    = {Shrestha, S. and Irby, E. and Thapa, R. and Das, S.},
  title     = {SoK: A systematic literature review of Bluetooth security threats and mitigation measures},
  booktitle = {International Symposium on Emerging Information Security and Applications},
  pages     = {108--127},
  year      = {2022}
}

@inproceedings{b22,
  author    = {Spill, D. and Bittau, A.},
  title     = {BlueSniff: Eve meets Alice and Bluetooth},
  booktitle = {Proceedings of the USENIX Workshop on Offensive Technologies (WOOT)},
  pages     = {1--10},
  year      = {2007}
}

@misc{b23,
  author       = {{Bluetooth SIG}},
  title        = {Bluetooth Core Specification},
  howpublished = {Version 5.1, Volumes 0--6},
  year         = {2019},
  month        = {January}
}

@misc{ieee_ra,
  title        = {{Guidelines for Use of Extended Unique Identifier (EUI), Organizationally Unique Identifier (OUI), and Company ID}},
  author       = {{IEEE Registration Authority}},
  year         = {2023},
  url          = {https://standards.ieee.org/products-programs/regauth/}
}

@inproceedings{sobot2022machine,
  title={Machine learning methods for device identification using wireless fingerprinting},
  author={Sobot, Srdjan and Ninkovic, Vukan and Vukobratovic, Dejan and Pavlovic, Milan and Radovanovic, Milos},
  booktitle={2022 International Balkan Conference on Communications and Networking (BalkanCom)},
  pages={183--188},
  year={2022},
  organization={IEEE}
}

@misc{bezawada2018iotsense,
      title={IoTSense: Behavioral Fingerprinting of IoT Devices}, 
      author={Bruhadeshwar Bezawada and Maalvika Bachani and Jordan Peterson and Hossein Shirazi and Indrakshi Ray and Indrajit Ray},
      year={2018},
      eprint={1804.03852},
      archivePrefix={arXiv},
      primaryClass={cs.CR},
      url={https://arxiv.org/abs/1804.03852}, 
}

@inproceedings{Celosia2019SavingPrivateAddresses,
  author    = {Guillaume Celosia and Mathieu Cunche},
  title     = {Saving Private Addresses: An Analysis of Privacy Issues in the Bluetooth-Low-Energy Advertising Mechanism},
  booktitle = {Proceedings of the 16th EAI International Conference on Mobile and Ubiquitous Systems: Computing, Networking and Services},
  series    = {MobiQuitous},
  address   = {Houston, TX, USA},
  publisher = {Association for Computing Machinery},
  year      = {2019},
  month     = nov,
  numpages  = {10},
  isbn      = {978-1-4503-7283-1},
  doi       = {10.1145/3360774.3360777}
}

@article{Celosia2020DiscontinuedPrivacy,
  author  = {Guillaume Celosia and Mathieu Cunche},
  title   = {Discontinued Privacy: Personal Data Leaks in Apple Bluetooth-Low-Energy Continuity Protocols},
  journal = {Proceedings on Privacy Enhancing Technologies},
  year    = {2020},
  volume  = {2020},
  number  = {1},
  pages   = {26--46},
  doi     = {10.2478/popets-2020-0003}
}

\begin{appendix}
    \section{Ethics}

This work involves the collection and analysis of BLE advertising traffic, which may include data from nearby third-party devices. Although such data is publicly broadcast and does not require pairing or user interaction, it may still raise privacy concerns, as device-level information could potentially be linked to individuals. To mitigate this risk, we will not release raw traces publicly; access to the dataset will be granted only upon request, while code for analysis will be made openly available. Furthermore, the techniques presented in this work could be misused to enable device tracking. Our goal is to highlight limitations of existing privacy mechanisms and contribute to their improvement, rather than to facilitate misuse.
\end{appendix}

\end{document}